# Applying l-Diversity in anonymizing collaborative social network

G.K.Panda
Department of CSE & IT
MITS, Sriram Vihar
Rayagada, INDIA
gkpmail@sify.com

A. Mitra
Department of CSE & IT
MITS, Sriram Vihar
Rayagada, INDIA
mitra.anirban@gmail.com

Ajay Prasad
Department of CSE
Sir Padampat
Singhania University,
Udaipur, INDIA
ajayprasadv@gmail.com

Arjun Singh
Department of CSE
Sir Padampat
Singhania University,
Udaipur, INDIA
vitarjun@gmail.com

Deepak Gour
Department of CSE
Sir Padampat
Singhania University,
Udaipur, INDIA
deepak.gour@spsu.ac.in

*Abstract*— To date publish of a giant social network jointly from different parties is an easier collaborative approach. Agencies and researchers who collect such social network data often have a compelling interest in allowing others to analyze the data. In many cases the data describes relationships that are private and sharing the data in full can result in unacceptable disclosures. Thus, preserving privacy without revealing sensitive information in the social network is a serious concern. Recent developments for preserving privacy using anonymization techniques are focused on relational data only. Preserving privacy in social networks against neighborhood attacks is an initiation which uses the definition of privacy called k-anonymity. k-anonymous social network still may leak privacy under the cases of homogeneity and background knowledge attacks. To overcome, we find a place to use a new practical and efficient definition of privacy called l-diversity. In this paper, we take a step further on preserving privacy in collaborative social network data with algorithms and analyze the effect on the utility of the data for social network analysis.

## I. INTRODUCTION

As the ability to collect and store more and more information about every single action in life has grown, huge amounts of details about individuals are now recorded in database systems. Social networks have always existed in society in varying forms. The record keeping power of computers and the advancement of internet, both the interactions within and scale of these social networks are becoming apparent. Individual social networks have proved fruitful and have been the topic of much research. However with the development of agencies, facilitators and researchers, the ability and desire to use multiple social networks collaboratively has emerged.

This has both positive and negative effects. The positive effects focus on many possibilities for enriching people's lives through new and improved social services and a greater knowledge of people's preferences and desires. The negative effect focuses on the concerns that private aspects of personal lives can be damaging if widely publicized. For example, knowledge of a person's locations, along with his preferences can enable a variety of useful location-based services, but public disclosure of his movements over time can have serious consequences for his privacy.

However, agencies and researchers who collect such data are often faced with a choice between two undesirable outcomes. They can publish personal data for all to analyze, this analysis may create severe privacy threats, or they can withhold data because of privacy concerns, this makes further analysis impossible and may hamper the social feel of the network and may lead to unpopularity of the site. Thus retaining individual privacy is really a concern to the social network analysis society.

### A. Need of privacy in Social Network data

Let us ponder on two examples of social sharing that lead to troublesome situations.

**Example 1**: The Enron corporation bankruptcy in 2001 made available of 500,000 email messages public through the legal proceedings and analyzed by researchers [7]. This data set has greatly aided research on email correspondence, organizational structure, and social network analysis, but it also has likely resulted in substantial privacy violations for individuals involved.

**Example 2:** Network logs are one of the most fundamental resources to any computer networking security professionals and widely scrutinized in government and private industry. Researchers analyze internet topology, internet traffic and routing properties using network traces that can now be collected at line speeds at the gateways of institutions and by ISPs. These traces represent a social network where the entities are internet hosts and the existence of communication between hosts constitutes a relationship. Network traces (even with packet content removed) contain sensitive information because it is often possible to associate individuals with the hosts they use, and because traces contain information about web sites visited, and time stamps which indicate periods of activity.

There are five basic types of IP address anonymization algorithms in use. These are: black-marker anonymization, random permutations, truncation, pseudonymization and prefix-preserving pseudonymization. Somehow these are trivial and there are certain mapping methods which still have a chance to get exposed on attacks. To eliminate the hurdle of sharing logs, strong and efficient anonymization techniques are very much essential. [7].

Recent work has focused on managing the balance between privacy and utility in data publishing, but limited to relational datasets. The definitions of k-anonymity [10] and its variants [2, 6] are promising. These techniques are commonly implemented with relational micro-data. While useful for census databases and some medical information, these techniques cannot address the fundamental challenge of managing social network datasets.

Bin Zhou and Jian Pei [15] proposed a privacy preservation scheme which deals against neighborhood attacks of social network using the definition of privacy called k-anonymity. However, k-anonymous social network still may leak privacy under the cases of homogeneity attacks and background knowledge attacks.

*B. Contributions and Paper Outline*

We propose an algorithm for the collaborative social network anonymization which can be extended to the higher level of security threat, where the adversary can have the information even about the vertices which are not the immediate neighbours of target vertex.

The basic definitions are provided in the next section. Section 3 describes the existing system to deal with the security in social network with k-anonymity as proposed by Zhou and Pei [15]. We extend the work to 2-neighborhood with an algorithmic approach and highlight possibilities of attacks. To overcome from such attacks in social network, we use a new practical and efficient definition of privacy called l-diversity [6]. It is proved [12] that l-diversity always guarantees stronger privacy preservation than k-anonymity. Section 4 highlights the proposed system of collaborative social network anonymization with equivalence relations and l-diversity method. Our goal is to enable the useful analysis of social network data while protecting the privacy of individuals. Finally section 5 gives the conclusion

II. DEFINITIONS AND NOTATIONS

In this section we reintroduce some basic notations that will be used in the remainder of the paper.

**Definition-1** (Modelling Social Network) A social network can be modelled as a simple graph, $G = (V, E, L, \xi)$ where, V is the set of vertices of the graph, E is the edge set, L is the label set and $\xi$ is the labelling function from vertex set V to label set L, $\xi = V \rightarrow L$.

**Definition-2** (k-Anonymity) A table T satisfies k-anonymity if for every tuple $t \in T$ there exists k-1 other tuples $t_{i1}, t_{i2}, ..., t_{ik-1} \in T$ such that $t[C] = t_{i1}[C] = t_{i2}[C] = ... = t_{ik-1}[C]$ for all $C \in QI$

**Theorem 1** (k-Anonymity): Let G be a social network and G' an anonymization of G. If G' is k-anonymous, then with the neighbourhood background knowledge, any vertex in G cannot be re-identified in G' with confidence larger than 1/k.

**Definition-3** (Naive Anonymization) The naive anonymization of a graph $G = (V, E)$ is an isomorphic graph, $G_{na} = (V_{na}, E_{na})$, defined by a random bijection $f: V \rightarrow V_{na}$. The edges of $G_{na}$ are $E_{na} = \{(f(x), f(x')) | (x, x') \in E\}$

**Definition-4** (Black-marker Anonymization) It replaces all IP addresses with a constant. It is quite similar to the affect as simply printing the log and blacking-out all IP addresses. This method is completely irreversible.

**Definition-5** (Random permutation Anonymization) This method creates a one-to-one correspondence between anonymized and unanonymized addresses that can be reversed by one who knows the permutation.

**Definition-6** (Truncation Anonymization) A fixed number of bits is decided upon (8, 16 or 24) and everything but those first bits are set to zero.

**Definition-7** (Pseudonymization) It is a type of anonymization that uses an injective mapping such as random permutations.

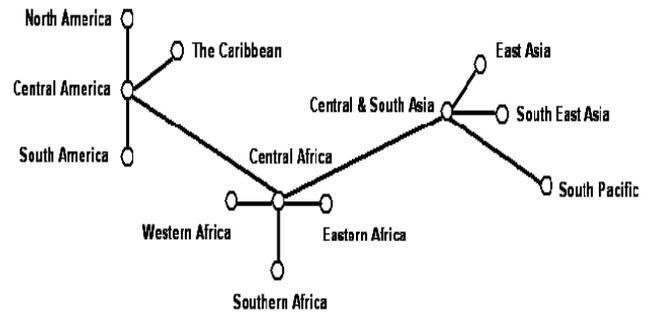

Figure 1. (a): A social network of interpol, G

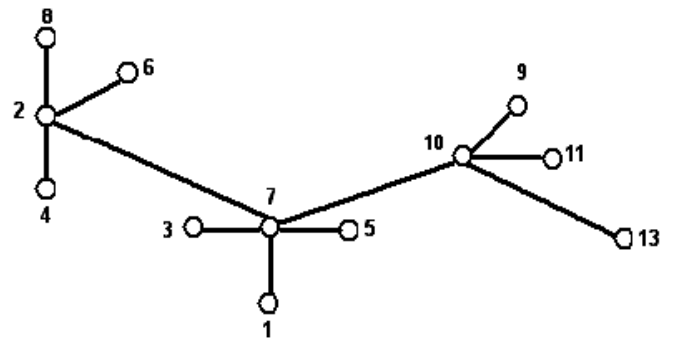

Figure 1. (b): The Naïve Anonymization of G

| North America | 8 |
|---|---|
| Central America | 2 |
| South America | 4 |
| The Caribbean | 6 |
| Central Africa | 7 |
| Western Africa | 3 |
| Eastern Africa | 5 |
| Southern Africa | 1 |
| Central & South Asia | 10 |
| East Asia | 9 |
| South East Asia | 11 |
| South Pacific | 13 |

Figure 1. (c): The Anonymization mapping

**Definition-8** (Prefix-preserving Anonymization) In this anonymization, IP addresses are mapped to pseudo-random anonymized IP addresses by a function where $\forall 1 \leq n \leq 32$, Pn(x) = Pn(y) if and only if Pn($\tau$(x))=Pn($\tau$(y)).

### III. THE EXISTING SYSTEM

The naive anonymization of a social network is to publish a version of the data that removes identification attributes. In order to preserve node identity in the graph of relationships, synthetic identifiers are used to replace them. *Figure 1* represents a social network of international police organization (Interpol), their naive anonymization and the anonymization mapping.

#### A. Publishing Social Network Data

To date publish of a giant social network jointly from different parties is an easier collaborative approach. Agencies and researchers who collect such social network data often have a compelling interest in allowing others to analyze the data. In many cases the data describes relationships that are private and sharing the data in full can result in unacceptable disclosures. Thus, preserving privacy without revealing sensitive information in the social network is a serious concern. Recent developments for preserving privacy using anonymization techniques are focused on relational data only

#### B. Preserving privacy in Social Networks using k-anonymity

The algorithm suggested by Bin and Jian [15] to anonymize a social network describes two basic steps as summarized below.

Step-1 Neighborhood Extraction and Vertex Organization

The neighborhood of each vertex is extracted and different components are separated. As the requirement is to anonymize all graphs in the same group to a single graph, isomorphism tests are conducted. For this purpose, for every component of the vertex the following steps are performed. Firstly all possible DFS trees are constructed for the component. Next, its DFS codes are obtained with respect to every DFS tree. Further the minimum DFS code is selected. This code is said to represent the component. Minimum DFS code has a nice property [14]: two graphs G and G0 are isomorphic if and only if DFS(G) = DFS(G0). Then neighborhood component code order is used to obtain single code for 1 vertex.

Step-2 Anonymization

Anonymization is done by taking the vertices from the same group. If the match is not found, the cost factor is used to decide the pair of vertices to be constructed.

Algorithm for k-Anonymization of one neighborhood

Input: A social network G=(V, E), the anonymization requirement parameter k, the cost function $\alpha, \beta$ and $\gamma$ ;

Output: An anomyzed graph G';

    1: initialize G' = G;

    2: mark $v_i \in V(G)$ as "unanonymized";

    3: sort $v_i \in V(G)$ as VertexList in neighbourhood size - descending order;

    4: WHILE (VertexList ≠ $\phi$ ) DO

    5: let SeedVertex = VertexList.head() and remove it from VertexList;

    6: FOR each $v_i \in$ VertexList DO

    7: calculate Cost(SeedVertex $v_i$) using the anonymization method for two vertices;

    END FOR

    8: IF (VertexList.size() ≥ 2k - 1) DO

    let CandidateSet contain the top k - 1 vertices with the smallest Cost;

    9: ELSE

    10: let CandidateSet contain the remaining unanonymized vertices;

    11: suppose CandidateSet= {u1,...um} anonymize

    Neighbour(SeedVertex) and Neighbour(u1)

    12: FOR j = 2 to m DO

    13: anonymize Neighbour(uj) and {Neighbour(SeedVertex), Neighbour(u1) ....... Neighbour(uj-1)}mark them as "anonymized";

    14: update VertexList;

    END FOR

    END WHILE

#### C. Algorithm for k-Anonymization of two neighbourhoods

**Step-1:** Let u, v∈ V(G), u and v have similar neighbourhoods. Then, the labels are generalized or left unchanged, so that the neighbourhoods of u and v are isomorphic. Also, the labels of the vertices are same in both the neighbourhoods.

**Step-2:** If the neighbourhoods are not similar, the cost is calculated and the pair of vertices with the minimum cost is considered.

**Step-3:** The edges necessary are added to make them similar.

**Step-4:** The process of Step-1 is applied on the vertices pair.

*D. Possible attacks on k- Anonymity*

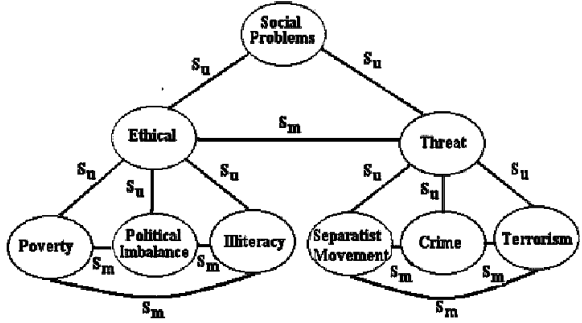

Figure 2. A social Network

Followings are two types of attacks that disclose sensitive information in k-anonymity [l-diversity] under two cases. First, if an attacker can discover the values of sensitive attributes when there is little diversity in those sensitive attributes. Second, k-anonymity does not guarantee privacy against attackers using background knowledge (Attackers often have background knowledge).

Since both of these attacks are plausible in real life, it is required to go forward with the new definition of privacy that takes care of diversity and background knowledge.

*Homogeneity Attack:* k-Anonymity can create groups that leak information due to lack of diversity in the sensitive attribute.

*Background knowledge attack:* k-Anonymity does not protect against attacks based on background knowledge.

IV. THE PROPOSED SYSTEM

In recent years, motivated by quality-aware scenarios like imprecise observations, there has been a growth of interest in models and algorithms for handling uncertain data, i.e. data describing many alternatives. Several working models of uncertain data have been proposed, which precisely describe many possible worlds by outlining the alternatives for possible events and the correlations or in dependencies between them. Given data presented in such models, there has been much effort in studying how to efficiently evaluate queries and perform analysis over the uncertain data and come up with a compact description of the possible answers to the queries.

It is quite clear to observe that, there is an important connection between the topics of Uncertain Data and Data Anonymization. The process of data anonymization introduces uncertainty into data that was initially certain. Data anonymization provides with principles methods for query evaluation and Uncertainty deals with a natural application area for uncertain data and both with a rich set of challenging problems.

*A. Anonymity with structural equivalence*

Nodes that look structurally similar may be indistinguishable to an adversary, in spite of external information.

**Definition-3:** In a social network, a pair of nodes x and y are said to be structurally equivalent [3] ( $x \approx y$ ) when

1. $V(x, y) = \wedge (x, y)$

2. for any $v \in \wedge (x, y)$, $R(x, v) = R(y, v)$

3. if $y \in G(x)$, $R(x, y) = R(y, x)$

The equivalence ( $x \approx y$ ) certainly means that x and y share a common set of relationships with a particular group of other nodes. It may not be necessary that x and y are to be directly connected/related.

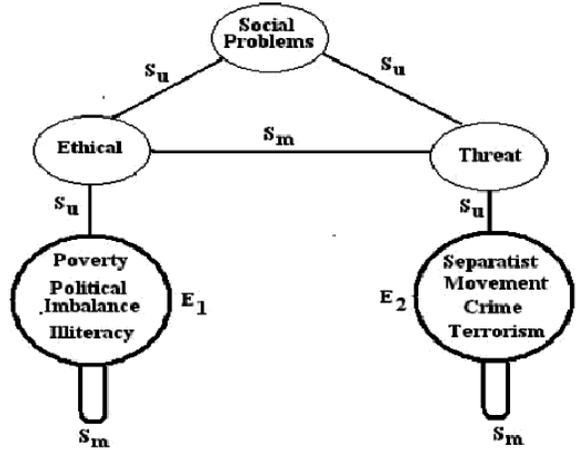

Figure 3. Reduction Social Network

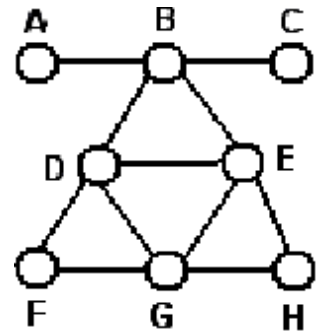

Figure 4. A Social Network Graph

**Theorem 2:** The equivalence relation induces a partition of N into disjoint equivalence classes satisfying:

1. x and y belong to the same equivalence class if and only if ( $x \approx y$ )

2. for $y \in G(x)$, R(x,y) is uniquely determined by the equivalence classes of x and y.

The key issue of this theorem is that the multitude of relations in the network can be summarized by simply observing the relations that exist among equivalence classes. *Figure 2* shows a social network with a set of investigated social problems like Ethical Problems, Threats, Poverty, Political imbalance, Illiteracy, Separatist movement, Crime, Terrorism. *Figure 3* shows the reduction social networking with structural equivalence [3].

*B. Automorphic Equivalence*

**Definition-9** (Automorphic equivalence): It is one of strong form of structural similarity between nodes. Two nodes x, y ∈ V are automorphically equivalent (≈A) if there exists an isomorphism from the graph onto itself that maps x to y.

Automorphic equivalence induces a partitioning on V into sets whose members have identical structural properties. An adversary even with exhaustive knowledge of a target node's structural position, cannot isolate an individual beyond the set of entities to which it is automorphically equivalent. These nodes are structurally indistinguishable and observe that nodes in the graph achieve anonymity by being "hidden in the crowd" of its automorphic class members.

*C. Vertex refinement*

This technique originally developed to efficiently test for the existence of graph isomorphism. Here, the weakest knowledge query, H0 simply returns the label of the node. The queries are successively more descriptive, like $H_1(x)$ returns the degree of x, $H_2(x)$ returns the list of each neighbours' degree and so on. In general $H_i(x)$ returns the multiple set of values which are the result of evaluating $H_{i-1}$ on the set of nodes adjacent to x: $H_i(x) = \{H_{i-1}(z_1), H_{i-1}(z_2),..., H_{i-1}(z_m)\}$ where $z_1,...z_m$ are the nodes adjacent to x. Two nodes x, y in a graph are equivalent relative to $H_i$ denoted $x \cong H_i$, if and only if $H_i(x) = H_i(y)$. Figure 4 is a simple social network graph, Table 1 represents the vertex refinement table and Table 2 is the equivalence class need to be further anonymized.

The proposed algorithm to anonymize a collaborated social network with l-diversity describes with three basic steps as summarized below.

Step-1 Formation of the collaborative social network by adding individual social networks

Step-2 Generating equivalence classes of nodes of the social network with vertex refinement having automorphic structural equivalence.

Step-3 Anonymization using l-diversity principle

*D. The l-diversity Anonymization*

**Definition 10:** (The l-diversity principle) A q*-block is l-diverse if contains at least l "well-represented" values for the sensitive attribute S. A table is l-diverse if every q*-block is l-diverse.

**Definition 11:** (The l-diversity principle in social network) An equivalence class of social network node implied by vertex refinement with structural equivalence is said to have l-diversity if there are at least l "well-represented" values for the sensitive node. It is said to have l-diversity if every equivalence class has l-diversity.

*E. Algorithm for Collaborative Social network with l-diversity*

**Input:** A social network G = (V, E), l-diversity parameter

**Output:** An anonymized graph

1: S- The collaborative social network

2: S(o, e)

    o     – A node within the social network n

    $o_j$    – The set of social networks that have provided with attributes, the numbering is made at the individual attribute level.

    e     - A set of edges related to node o

    $e_i$    – The set of social network from a user query

    Sr    – The resulting social network from a user query

3: N – A social network is being added to S

   N(d, g) -

       d- A node within the social network N

       g- A set of edges related to node o

4: R- A revocation social network being removed from S,

   R(d, g)

       d- A node within the social network R

       g- A set of edges related to node o

5: U- A user of S

   Uq – A query containing attributes to look for within S

   a(0)– An attribute or set of attributes of a node which can be used to uniquely identify the node

1: FOR each N(di, g)

2: IF ( a(S(o)) = = a(N(d))) THEN

3:  FOR each attribute within d

4:  IF the attribute matches in o

5:   $o_j = o_j + N$

5:  ELSE

6: ADD new attribute from d to o with $o_j = N$

7: END IF

8: FOR any edge within e

9: IF it matches an edge within g

10: $e_i = e_i + N$

11: ELSE

12: ADD non-matching edges within set g

    to set e with $e_i = N$

  END IF

  END FOR

13: Add new node and edge set $N(d_i, g)$ to S

  14: anonymize $N(d_i)$ and $N(g)$

  END IF

  END FOR

## V. Conclusions

In this paper we tried to focus the important problem of preserving privacy in publishing collaborative social network data which is really an important concern. We have referred the k-anonymity algorithm for social network using one neighbourhood and extended to two neighbourhood approach. Further we have discussed the relevant attacks to k-anonymity. We have studied the extent to which structural properties of a node using equivalence can serve as a basis of re-identification in anonymized social networks. This work further has potentiality to extend using the rich rough set theory. A l-diversity social network still may leak privacy. An adversary may have some prior belief about the sensitive attribute value of an individual before seeing the released table. After seeing the released table, the adversary may have a posterior belief. Information gain i,e., the difference between the posterior belief and the prior belief is the factor to leak privacy. Some mechanism analogous to t-closeness should be introduced.